\documentclass[12pt]{article}

\usepackage{enotez}
\usepackage[utf8]{inputenc}
\usepackage[bottom = 1.5in, top = 1.5in, total={6.6in, 9in}]{geometry}
\usepackage{mathptmx}
\usepackage{graphicx}
\usepackage{epigraph}
\usepackage{comment}
\usepackage[table]{xcolor}

\usepackage{amsfonts,amsmath,amssymb,amsthm,mathrsfs,url,hyperref, graphicx, pxfonts, xcolor}

\usepackage{physics}
\usepackage{microtype}
 \usepackage[labelfont=bf]{caption}
\usepackage{multirow}
\usepackage{amsmath}
\usepackage{amssymb}
\usepackage{cleveref}
\usepackage{ragged2e}
\usepackage{natbib}
\newcommand{\x}[1]{{\color{black}#1}}
\newcommand{\y}[1]{{\color{black}#1}}
\newcommand{\w}[1]{{\color{black}#1}}
\newcommand{\z}[1]{{\color{black}#1}}
	
	\title{Physical Coherence and Time's Emergence}
	\author{Eugene Y. S. Chua \\ Nanyang Technological University Singapore \\ eugene.chuays@ntu.edu.sg}
	\date{Accepted at \textit{Synthese}. Preprint of 24 May 2026.}
	
\begin{document}
		
		\maketitle
		
		\begin{abstract}
			\noindent It is often said that time vanishes in quantum gravity. One general approach to quantum gravity accepts this fundamental timelessness but seeks to derive time's emergence at a non-fundamental level. To better assess such approaches, I develop the criterion of physical coherence and situate it in context by applying it to two programs for time's emergence, drawing from recent works by Chua and Callender (2021) and Chua (2025): semiclassical time and thermal time. Unlike some recent arguments for the metaphysical incoherence of time's emergence, which rule out all claims of time’s emergence `from on high' once we’ve fixed a definition of metaphysical emergence, my criterion of physical coherence leaves open the possibility that some programs in quantum gravity may succeed on their own terms in providing a physically coherent derivation of time from no-time. This sets a challenge for proponents of time's emergence to clarify the conceptual foundations of their program, while at the same time acting as a litmus test for a program's success.
		\end{abstract}
		
		\section{Introduction}
		
		In quantum gravity, \z{there are serious proposals that suggest that space, time, or both, vanish at the fundamental level. For instance, Misner, Thorne, and Wheeler (1973, 1183) write that considerations about quantum physics lead one to conclude that ``there is no spacetime, no time, there is no before, there is no after. The question of what happens ``next'' is without meaning.''} In canonical quantum gravity (CQG), one is led, upon imposition of certain constraints and standard quantization procedures onto general relativity,\footnote{\x{See Dirac (1964) for the classical treatment of `Dirac quantization', and Kuchař (1993) and Mozota Frauca (2023b) for discussion of how it applies to general relativity. Kuchař (1993), Mozota Frauca (2023b), and Gryb and Th\'ebault (2016) also raise worries for whether Dirac's quantization procedure is the right way to quantize general relativity. Here, I assume that the standard approach to quantization is correct.}} to the so-called Wheeler-DeWitt equation,
		\begin{equation} \label{eq:1}
			\hat{H}|\Psi\rangle = 0,
		\end{equation}
		which appears absent of dynamical evolution. Here, $\hat{H}$ is the Hamiltonian constraint operator, and $|\Psi\rangle$ is associated with the wave-functional $\Psi$ representing both matter content and geometry.\footnote{More concretely, we find an infinite set of Hamiltonian constraints of the form:
			\begin{equation*}\label{hamcon}
				\hat{H}(f) = \int d^3x \hat{H}(x) f(x)
			\end{equation*}
			where $\hat{H}(x)$ is the Hamiltonian constraint density at each point of space, and $f(x)$ is a test function. \textit{All} these constraints must vanish, i.e. something like \eqref{eq:1} holds for all these constraints so that \textit{any} physical Hamiltonian vanishes.} While the Schrödinger equation,
		\begin{equation} \label{eq:2}
			\hat{H}|\Psi\rangle = i\hslash \frac{\partial |\Psi\rangle}{\partial t}\; ,
		\end{equation}
		describes the time-evolution of $|\Psi\rangle$, the Wheeler-DeWitt equation doesn't. Because the $t$ parameter vanishes, it is said that time itself `disappears' and that the fundamental ontology of CQG is timeless, containing no reference to time or concepts depending on time. 
		
		Yet, it seems manifest that our familiar physical systems are  -- or at least appear to be -- evolving \textit{in time}. Any theory that tells us otherwise must provide a satisfactory story for why familiar physical systems (appear to) evolve in time. This gives rise to the problem of recovering time-evolution from fundamentally timeless ontology: the \textit{problem of time}.\footnote{Though often branded as \textit{the} problem of time, this is actually just \textit{one} of many problems of time. See Thébault (2021) for a philosophical overview of the problem of time. For a full overview of the numerous problems categorized as `problems of time' in CQG, see Anderson (2017). See also Kuchar (1991), Isham (1993), and Kuchař (2011).} In Chua and Callender's (2021) words, it is the problem of recovering `time from no-time'. 
		
		There are broadly three sorts of responses one might adopt. First, one can be a \textit{fundamentalist}. On this view, time -- where it exists -- is fundamentally there, albeit `hidden'. Even though the $t$ parameter vanishes in the Wheeler-DeWitt equation, one seeks to find another parameter with which one can define dynamical equations like \eqref{eq:2}.\footnote{See e.g. Greensite (1990), Padmanabhan (1990), Callender and Weingard (1994), Shtanov (1996), Vink (1992), recently Pinto-Neto and Struyve (2019).} Second, one can be an \textit{eliminativist}. On this view, time doesn't exist fundamentally; one abolishes the concept and accepts that what we call `time' is illusory in some sense.\footnote{Its most prominent defender is Barbour (1999), who recasts physics into timeless form and argues that all that exists are spatial configurations of matter. See e.g. Barbour (2012), and also Baron, Miller and Tallant (2022).}
		
		My focus is instead on what I'll call the \textit{middle way}:\footnote{\y{I emphasize that this approach is not novel nor original to my discussion here. Indeed, as an anonymous reviewer points out, it's the ``standard position among philosophers and physicists".}} on this view, one accepts, as eliminativists do, that \textit{time does not fundamentally exist} in a quantum-gravitational universe. However, contrary to eliminativists, the middle way also accepts, as fundamentalists do, that time exists, though (unlike fundamentalism) only at some \textit{non-fundamental} level. Furthermore, the middle way typically involves an \textit{explanatory component}: one seeks not only to \textit{claim}, brutely, time's emergence from timelessness, but also to \textit{explain} how, through physics, time emerges. 
		
		Recently, Baron, Miller and Tallant (2022) have argued that the middle way runs into \textit{metaphysical incoherence}: \w{standard metaphysical accounts of emergence in terms of general relations between more fundamental and less fundamental objects such as composition, constitution, or realization (Baron, Miller and Tallant 2022, \S6), arguably rely on spatiotemporal properties which don't exist in a fundamentally non-spatiotemporal universe. For instance, the standard notion of composition in terms of parthood (that A composes B when A is a part of B arranged in the right way) seems to depend, at least, on ``a fairly minimal connection between the location of the whole and the location of the parts", but this means that ``it is difficult to make sense of the idea that spacetime or spatiotemporally located entities are composed of entities that are not spatiotemporal." (127) Similarly, they argue, for constitution and emergence. However, since all known accounts of emergence employ such spatiotemporal properties, the metaphysics of the middle way appears to be metaphysically incoherent: it leads to circularity.}
		
		While I sympathize with their critique, metaphysical worries, \textit{external} to the physical project of deriving time from no-time, may not hold sway over physicists and naturalistically-inclined philosophers of physics since these metaphysical worries tend to depend little on the physical details of any particular quantum gravity program. Having fixed a definition of metaphysical emergence, we rule out \textit{all} claims of spacetime emergence by fiat. 

        My concern here is different. I ask whether a proposal for recovering time is \emph{physically coherent} on its own terms. A program is physically coherent, in the sense relevant here, only if the formal tools used to derive effective temporal structure can themselves be given a physical interpretation consistent with the timeless starting point of the theory. Metaphysical incoherence and physical incoherence may therefore come apart. One can take the metaphysical worry seriously while still asking the more local question of whether a given quantum-gravity program has earned its own emergence claim.
		
		In this paper, I propose an internal criterion of (in)coherence distinct from metaphysical incoherence -- \textit{physical (in)coherence} -- which leaves open the possibility that some programs for the middle way might succeed \textit{on their own terms}, as opposed to whether a metaphysical account of emergence exists `from on high'. In \S2, I develop the criterion of physical coherence and argue that the middle way may fail to satisfy this new criterion. Specifically, I argue that the explanatory project of the middle way -- in trying to derive time from no-time using physics -- may run into incoherence if the formal tools employed in such derivations require dynamics and time for their interpretation, implicitly or explicitly. By briefly considering parts of the physical reasoning \textit{internal} to two concrete proposals for the middle way, semiclassical time (\S3.1) and thermal time in \S3.3, I show how one may apply this standard of physical coherence in context, discussing Huggett and Th\'ebault's recent criticisms of Chua and Callender's analysis along the way in \S3.2.
		
		To emphasize, I am \textit{not} claiming that \textit{all} programs in quantum gravity purporting to derive time's emergence must, of necessity, run into physical incoherence. I leave it open that some proposal for the middle way may succeed. This points to the fruitfulness of analyses of physical coherence:\footnote{Some recent works in this vein can be found in Mozota Frauca (2023a, 2024).} it presents a challenge, for followers of the middle way to clarify the physical and metaphysical status of their programs, a road-map, for future work in search of similar problems in other branches of quantum gravity where proposals for the middle way are proposed, and a litmus test, for gauging their success or failure. 

        \w{It must be noted as well that physical coherence is unlikely to be sufficient for the successful emergence of time. A program may avoid internal circularity in its use of approximations and still fail for other reasons: the recovered temporal structure may be too weak, too idealized, empirically inadequate, or metaphysically obscure. However, I do think that passing the test of physical coherence is best understood as a necessary condition on successful accounts of time's emergence. If an account does not possess physical coherence, it cannot begin to be a physical story for how time emerges. If a proposal cannot supply a physically coherent interpretation of the steps by which time is recovered, no matter how formally consistent, then it has not yet earned the stronger conclusion that time emerges from no-time.}

        \w{Finally, this criterion is intended to complement broader work on spacetime emergence in quantum gravity, such as those accounts discussed in W\"uthrich and Huggett (2025). In particular, recent spacetime-functionalist treatments ask how non-spatiotemporal structures can realize spacetime roles, which invites metaphysical questions about the nature of realization. The question pursued here is a more local and contextual one: when a \textit{specific} program invokes approximations, ans\"atze, or idealizations to recover a functionally defined emergent time from a fundamentally timeless ontology, can those intermediary steps themselves be given a physical interpretation consistent with the program's timeless starting point? Physical coherence is thus a constraint on particular derivations within broader emergence programs, including functionalist ones.
        
        As we'll see later in \S2, group field theory, discussed by Huggett (2022) as a case of spacetime emergence, may face similar worries of physical incoherence. More generally, I believe that the same diagnostic question can be asked (and may be satisfied or otherwise) by any theory purporting to recover time from no-time: do the relevant intermediary tools presuppose the very spatiotemporal structure they are meant to explain?}
		
		\section{The physical (in)coherence of the middle way}

        \subsection{Physical coherence}
		
		The standard metaphysics of emergence seems to require spacetime in its application, and so cannot be applied to non-spatiotemporal ontology without risking metaphysical incoherence. This has led some, such as Baron, Miller and Tallant (2022), to raise worries about the very possibility of spacetime emergence as a matter of metaphysics. For supporters of the middle way, however, one might retort: so much the worse for metaphysics. Physicists working on the middle way are unlikely to be worried by esoteric concerns about the metaphysics of emergence, e.g., mereology or the nature of composition. These are questions \textit{external} to the practice of physics. Furthermore, metaphysical criticisms of spacetime emergence appear to be all-or-nothing: \textit{all} claims of spacetime emergence -- \textit{all} quantum gravity programs walking the middle way -- are metaphysically incoherent if they are right. As McKenzie (2020) forcefully argues, metaphysical claims such as these tend to have a ``dichotomous, crude, character", which ``undermines the ability of our metaphysical claims to participate in the progress of science" (21). 
		
		I am inclined to agree. If spacetime emergence is metaphysically incoherent `from on high', then \textit{all} programs in quantum gravity purporting to walk the middle way are necessarily rendered defunct once we've fixed a metaphysical definition of emergence.\footnote{I note that Baron, Miller, and Tallant (2022, 151--152) do leave open the possibility that an account of emergence arises which can avoid metaphysical incoherence. For instance, they suggest, as possible research directions, relations which are analogous to mereological relations or grounding-based accounts. Nevertheless, these remain \textit{metaphysical} accounts which do not engage with the subtleties of any particular quantum gravity program, and will still make blanket judgments on the viability of \textit{any} program without engaging with them on their own terms.} 
		
		In my view, a more fruitful criterion -- and one in line with the practice of physics -- would be something that is sensitive to the details and subtleties of each program, can be satisfied by certain programs and not others, and can therefore provide an operable standard of success/failure for any particular program. 
		
		With that said, I am sympathetic to the worry of incoherence. In what follows I'll develop a physical analogue of metaphysical incoherence which may also apply to the middle way. In trying to explain how, via physics, time could emerge from timelessness, we must employ certain formal tools. I'll suggest that these tools may be justified, usually implicitly, by dynamical notions. In such cases, time's emergence might not only be metaphysically incoherent from some external perspective as a matter of necessity, but also \textit{physically incoherent} in light of considerations about how to interpret these tools \textit{internal} to physics practice. Crucially, physical incoherence, unlike metaphysical incoherence, leaves open the possibility that some programs for the middle way may succeed (or fail) \textit{on their own terms}, as opposed to `from on high'. 
		
		\y{So what does an analysis of physical coherence amount to, in the context of time's emergence? Every program seeking to walk the middle way has the same recipe, construed broadly in terms of \textit{formalism} and \textit{interpretation}. On the side of formalism, every program $(a)$ starts with a mathematical theory, e.g. a set of equations, $(b)$ introduces a set of \textit{formal tools}, e.g., a variety of ans\"atze, approximation techniques, idealized assumptions, and $(c)$ arrives at a new mathematical theory, a new set of equations. On the side of interpretation, either implicitly or explicitly, every program $(a')$ provides a physical interpretation of $(a)$ as fundamental and timeless, $(b')$ provides some physical interpretation for the formal tools, and $(c')$ provides an interpretation of $(c)$ in non-fundamental and temporal terms. 
			
		Given the recipe, a demand for physical coherence comes in two parts: first, $(a)$-$(c)$ must be \textit{formally consistent}. This is perhaps obvious; we want our derivation of time from timelessness to have no mathematical contradictions. Second, $(a')$-$(c')$ must be \textit{interpretively} consistent. Of course, both the starting point -- the fundamental timeless theory -- and the end point -- the non-fundamental temporal theory -- must be physically interpreted: the former will have, among other things, timeless ontology, while the latter will have temporal ontology. However, physical coherence \textit{also} demands \textit{interpretive consistency}: every formal tool employed along the way must be given an interpretation in terms of the fundamentally timeless ontology of the fundamental theory and lead to the appearance of non-fundamental temporal ontology, without circularity or contradiction. This, I think, is simply a clarification of the aforementioned explanatory demand of the middle way: what we want is not merely a brute claim of time's emergence, but to explain through physics \textit{why} time emerges from a timeless world.

        Physical coherence therefore allows for each program for the middle way to provide its own terms and conditions for interpreting their starting point, proposed formal tools, and end point, leaving it open that there can be some combination of interpretations of $(a')$-$(c')$ that can succeed by the program's own terms.

        The demand for physical coherence might seem prima facie trivial: after all, `providing a consistent physical interpretation' seems like a fairly weak desideratum. However, as will be clear from the GFT example below (and later discussion), this is actually a \textit{highly non-trivial} criterion for a given proposal in quantum gravity. Because of the inter-theoretic and highly technical nature of work in quantum gravity, theorists often borrow formal tools from many different physical contexts to develop their new proposals (such as phase transitions from condensed matter systems, in the case of GFT below). However, the intended physical interpretation of these formal tools both in the original context and the new quantum gravity context, in terms of physical systems in the world,  is not always clear. Sometimes it might be assumed that `something like' the original interpretation can be expected to hold in the new context without much justification. As a result, the physical incoherence of these proposals can wind up being masked in the technicalities, and the work of uncovering these issues is certainly non-trivial. Furthermore, it might be the case that `standard' interpretations of these formal tools, applied to the context of quantum gravity, imply physical incoherence, but that there exist better interpretations which avoid the worry. In such cases, we may obtain a clearer understanding of the program and why it can walk the middle way after all. In short, the evaluation of a proposal's physical (in)coherence is hard work, but it is work that can be (and have been) done by philosophers of physics.}
		
		A quick clarification: `physical' here does \textit{not} necessarily mean `in temporal terms', though it is certainly true that most of our familiar physical systems are, in fact, temporally situated systems, and our principles and procedures -- such as those concerning state-to-state transitions -- are often couched in temporal terms. However, to provide a physical interpretation does require, at the very least, telling a story that would be consistent with the standards of physical practice, e.g., how these formal tools are interpreted and used for ordinary systems. And if the proposed interpretation deviates from this standard use, its proponent had better provide a satisfactory alternative interpretation for these formal tools. I don't think that it's worth arguing about `physicality' in the general sense. Rather, in what follows, I situate the standard of physical coherence in concrete contexts, and let context do the job of cashing out the notion of physicality for particular procedures and tools. 
                
        \x{To see that a demand for physical coherence can do work, we'll see in detail in \S3.2 that some philosophers and physicists may reject a demand for physical coherence as I have laid out here, because they reject the second criterion of physical coherence, that $(a')$-$(c')$ must be jointly and consistently physically interpreted. They hold the view that all that matters is the formal consistency of $(a)$-$(c)$, along with a consistent interpretation for $(a')$ and a separate, consistent, interpretation for $(c')$, \textit{without} insisting for a physical interpretation of $(b')$ that could bring us from $(a')$ to $(c')$. For instance, they might think that any physical argument from $(a')$ to $(c')$ via $(b')$ is merely `heuristic' and shouldn't be taken seriously or literally, and it is just a brute mathematical fact that $(a)$ entails $(c)$ via $(b)$ which does not need physical interpretation. 
			
        I agree that mathematical developments often outpace physical interpretation, especially in cutting edge physics, so that tools from other domains are often introduced to formally obtain $(c)$ from $(a)$ without a clear physical interpretation of said tools. The physical interpretation may be difficult to provide as well. Indeed, in the case of thermodynamics some questions about a consistent physical interpretation of the formal tools, bridging the mechanical and the thermodynamic physical description of systems, continue even today.\footnote{See Frigg and Werndl (2021) for this problem in statistical mechanics.} All this is to say that establishment of physical coherence is certainly non-trivial, relative to the establishment of mathematical consistency. 
			
        However, I fail to see this challenge as a philosophical defense against the criterion of physical coherence, especially a demand for interpreting (b) physically and checking its interpretive consistency with (a') and (c') -- it is only a reason for why physical coherence takes} \x{time and effort to establish. The main worry is that the formal tools (b), as we'll see in the examples later, are often \textit{introduced} by hand as an ansatz or assumption only as a means to obtain (c), rather than something obviously compatible with and justified in their application to (a). Furthermore, the standard physical interpretations of (b) are often incompatible with bridging (a') and (c'). If we are allowed to assume anything, we are allowed to derive anything. The question is whether we are allowed to assume the application of some particular formal tool. The call for physical coherence, then, is simply a request to check that this assumption makes mathematical \textit{and} physical sense, and is not just putting in the physical phenomenon we want to derive in (c') by hand or in an ad hoc manner despite it missing in (a'), such as time or dynamical assumptions. Physical coherence is a demand that we provide a clear interpretation of the formal tools that are used to make claims of time's emergence from a timeless world -- that we explain, \textit{not just formally but physically}, why we are justified in using the formal tools to get us from $(a')$ to $(c')$. That is to say, I see the goal of establishing and assessing physical coherence as worthwhile, and non-trivial, as a matter of physics and philosophy.}

    \subsection{A warm-up: group field theory}
	\x{To see an assessment of physical coherence in action, consider group field theory (GFT) as a warm-up.\footnote{For reasons of space, I won't get into the details of GFT here. The analysis here is therefore necessarily a coarse-grained one, intended to emphasize how physical coherence is to be analyzed, rather than a final verdict on the physical coherence of GFT per se. See Mozota Frauca (2023a, 2024) for exposition and critique.} The formal starting point is $(a_{\text{GFT}})$ a group manifold; its physical interpretation is $(a'_{\text{GFT}})$ that the fundamental ontology comprises non-spatiotemporal group elements. One can define a Fock space over the group manifold such that generic states can be characterized as superpositions of excitations, just as in relativistic quantum mechanics or quantum field theory. The endpoint of this derivation is to obtain, firstly, $(c_{\text{GFT}})$ certain formally `well-behaved' Fock states which are highly coherent and symmetric. Secondly, these well-behaved Fock states are precisely those which can be $(c'_{\text{GFT}})$ interpreted as ``particles", ``atoms", or ``tetrahedra" of \textit{space} glued together to form something interpretable as our familiar relativistic 4-dimensional spacetime manifold. In these cases we may be tempted to say that (space-)time has emerged from the non-spatiotemporal group manifold, not only because of the mathematical fact that $(c_{\text{GFT}})$ can be formally derived from $(a_{\text{GFT}})$, but also the physical fact that $(c'_{\text{GFT}})$ can be interpreted in terms of $(a'_{\text{GFT}})$.
		
        However, as Mozota Frauca (2023a) discusses, generic Fock states are not interpretable in this nice geometric way of $(c'_{\text{GFT}})$. To argue that we should expect `nice' states rather than `problematic' ones, GFT employs $(b_{\text{GFT}})$ the formal tool of \textit{phase transitions} with broadly two ingredients:\footnote{ See Mozota Frauca (2023a, 2024).} (i) a formal approximation that the GFT Fock state can be described as a coherent and highly symmetric condensate state, i.e., every excitation is approximately in the same state and knowing the state of one excitation entails knowing the state of every excitation, and (ii) a formal analogy to standard condensed matter systems, with formally similar condensate states which can undergo thermodynamic phase transitions when extremizing some appropriate thermodynamic potential, e.g., the energy functional of the field theory associated with the condensate.
		
        (i) and (ii) motivate $(b'_{\text{GFT}})$ a particular physical interpretation of the formal tool of phase transitions, \x{\textit{as a genuine physical process}}: the idea that the world does undergo a phase transition from non-condensate and non-spatiotemporal states to condensate and spatiotemporal states -- just like ordinary Bose condensates. One argues that non-condensate states (generally uninterpretable geometrically) -- though generic -- might undergo phase transitions into condensate states (which \textit{can} be interpreted geometrically). Since condensate states, as a matter of formal fact, can be interpreted geometrically, specifically as smooth relativistic manifolds, proponents of a middle way in GFT conclude their task. Start with $(a_{\text{GFT}})$ and interpret it as timeless $(a'_{\text{GFT}})$. With the formal tool $(b_{\text{GFT}})$ interpreted as a genuine physical process $(b'_{\text{GFT}})$, we obtain the formal $(c_{\text{GFT}})$ interpreted as temporal $(c'_{\text{GFT}})$. 
		
        Mozota Frauca (2023a) has recently argued that this derivation of time from no-time in GFT runs into physical incoherence. Just to quickly sketch one criticism, in order to demonstrate the analysis of physical incoherence: one expects a physical interpretation of GFT phase transitions `akin to' ordinary phase transitions, but on standard physical interpretations of ordinary phase transitions -- in terms of the variation of thermodynamic variables e.g., temperature or pressure -- they occur \textit{over time}. This interpretation is not available for GFT states since those states are interpreted, per $(a'_{\text{GFT}})$, to be non-spatiotemporal and do not evolve over time in any meaningful sense. In Mozota Frauca's words: 
		\begin{quote}
			...in the case of a condensed matter system we can think of the system as embedded in a spatiotemporal structure in which the thermodynamic parameters change, while in the case of the theory of quantum gravity the theory itself defines a spatiotemporal structure, and introducing a further process of change seems to introduce an additional metatemporal dimension. (2023, 17)
		\end{quote}
		If he's correct, his analysis suggests that the middle way for GFT runs into physical incoherence: starting with a fundamentally timeless ontology, \textit{even if there were a formal or purely mathematical way of imposing these approximations related to phase transitions}, we cannot interpret these approximations in the intended way `just like' ordinary phase transitions without risking circularity, e.g., by assuming some additional temporal dimension. So we are never actually justified in arriving at the final temporal product of a spatiotemporal manifold. There is no consistent physical interpretation for all parts of the derivation, starting from the timeless ontology and ending in the final temporal product.}
		
		In what follows, I'll apply the criterion of physical coherence to two other programs walking the middle way, semiclassical time (\S3.1) and thermal time (\S3.3), responding to Huggett and Th\'ebault's (2025) (\textbf{HT}) recent criticisms of Chua and Callender's (\textbf{CC}) analysis of semiclassical time along the way in \S3.2. 

        \section{Physical coherence and two programs for the middle way}
		
		\subsection{Semiclassical time}
		
		One proposal for the middle way, discussed most prominently by Kiefer (2007), is the \textit{semiclassical time program}. Conceptually, the idea is simple: despite the fundamentally timeless ontology given by the Wheeler-DeWitt equation, time emerges in the non-fundamental, \textit{semiclassical}, level. This level is the regime in which the semiclassical approximations hold good. However, as I'll argue, the program faces the worry of physical incoherence once we scrutinize its use of the semiclassical lens.
		
		Remarkably, starting with $(a_{\text{SC}})$ the Wheeler-DeWitt equation \eqref{eq:1} where $\Psi$ is defined over both gravity and matter degrees of freedom and $(a'_{\text{SC}})$ can be interpreted as representing a timeless world, \textit{if} we are justified in using $(b_{\text{SC}})$ the semiclassical approximations, we can derive $(c_{\text{SC}})$ an equation formally resembling \eqref{eq:2} which $(c'_{\text{SC}})$ can be interpreted as governing the time-evolution of the wave-functional for the matter degrees of freedom $\phi$:\footnote{See Barbour (1993) for an elegant explanation, and Chua and Callender (2021) for discussion.}
		\begin{equation}\label{eq:3}
			i\hslash\frac{\partial}{\partial t}\psi(\phi,t;h_{ab})=\hat{H}^{m}(\phi;h_{ab})\psi(\phi,t;h_{ab})
		\end{equation}
		Here, $h_{ab}$ is the 3-metric associated with gravity, $\hat{H}^m$ is some Hamiltonian term which is interpreted as the energy of the matter part, while $\phi$ is a scalar field representing matter. Equation \eqref{eq:3} tells us how matter evolves over some `time'. However, this equation's `time' -- the semiclassical time -- is defined in terms of $h_{ab}$. The semicolons tell us that this matter wave-function is evaluated with respect to some gravitational $h_{ab}$. The crucial point: after we apply semiclassical approximations, gravity plays the role of a clock for matter, in order to define the notion of semiclassical time for the evolution of matter. Time emerges when the timeless world can be understood in a non-fundamental, semiclassical way.
		
		As Chua and Callender (2021) (\textbf{CC}) explain, there are three necessary steps to get from the fundamentally timeless universe to the semiclassical level where time emerges: starting with the timeless wave-function solution to the Wheeler-DeWitt equation (the fundamental timeless ontology), we assume a set of approximations (i.e. the formal tools) as follows: (1) we assume \textit{decoherence} in order to justify a particularly simple form of the wave-functional, (2) we assume the applicability of the \textit{Born-Oppenheimer approximation} to justify \textit{separating the gravitational degrees of freedom from the matter degrees of freedom}, and (3) we assume the applicability of the \textit{WKB approximation} in order to find a time parameter defined in terms of the gravitational degrees of freedom alone. These approximations let us formally derive \eqref{eq:3} (the temporal product), from the fundamentally timeless ontology. Each of these steps must then be interpreted appropriately to tell us a physical story of how fundamentally timeless wave-functionals lead to the appearance of some matter wave-function evolving over what seems like time. 
		
		However, as \textbf{CC} argues, the standard physical interpretation of these steps, i.e. a standard way to provide $(b'_{\text{SC}})$, a physical interpretation for the formal tools, appears to implicitly or explicitly require time in order to be applied. If this is the case, then we are not justified to apply these steps in a timeless world. Barring alternative interpretations (see \S3.2 for more discussion), the semiclassical time program appears to be \textit{physically incoherent}. Due to space constraints here, I cannot go into details as to how each step fails. Since all three steps are jointly necessary for time's emergence, all three steps must satisfy physical coherence. Hence, it suffices to show how step 1 -- the assumption of decoherence -- fails the test of physical coherence.\footnote{See Chua and Callender (2021) for details on the other two steps.} 
		
		\subsubsection{The time in decoherence}
		
		Without step 1, steps 2 and 3 do \textit{not} lead to the derivation of semiclassical time.\footnote{See Kuchař (1991) for discussion.}  Given step 1, even if the initial state of the universe is in an arbitrary superposition of states, decoherence suppresses interference between components of the reduced state in an appropriate basis, allowing one to treat the resulting branches as effectively non-interacting. If so, then even if the world were to be in some complicated superposition of states, we, living in one of these `branches' of the wave-function, can approximately neglect interference from other branches and treat the quantum state of our branch as an eigenstate in said basis, which is required for the derivation.\footnote{See Kiefer (2007, Ch. 10).} 
		
		The question is whether decoherence makes sense in a timeless world. Normally, decoherence is understood as a dynamic process, which is the source of \textbf{CC}'s concern. There are two main ways to see this worry. 
		
		First, according to Schlosshauer (2014), a quantum state of a subsystem decoheres when it interacts with \z{an environment}. For a subsystem $S$ in an arbitrary superposition of states
		\begin{equation}|S\rangle = \sum_{n}c_n |S_n\rangle\end{equation}
		interacting with an environment $E$, at some time $t$ after the interaction, the joint state of the environment and the system becomes:
		\begin{equation}|S+E\rangle = \sum_{n}c_n |S_n\rangle |E_n(t)\rangle\end{equation}
		For simplicity, consider $n = 2$. The density matrix describing the measurement outcomes on $S$, $\rho_S$, is:
		\begin{equation}
			\begin{split}
				\rho_S = \text{Tr}_E (\rho_{S+E}) & = \text{Tr}_E|S+E\rangle \langle S+E| \\
				& = |c_1|^2|S_1\rangle \langle S_1| + |c_2|^2|S_2\rangle \langle S_2| \\
				& + c_1 c_2^*|S_1\rangle \langle S_2| \langle  E_2(t)|E_1(t)\rangle + c_1^*c_2 |S_2\rangle \langle S_1| \langle E_1(t)|E_2(t)\rangle  
			\end{split}
		\end{equation}
		The last two terms, which I denote as $I$, represent the interference between the two superposed states and depend on $\langle E_1(t)|E_2(t)\rangle$ and $\langle E_2(t)|E_1(t)\rangle$. Generally, $\langle E_i(t)|E_j(t)\rangle$ indicates how orthogonal two states of the environment are. In standard models, these overlaps often decay approximately under Schrödinger evolution like        \begin{equation}\langle E_i(t)|E_j(t)\rangle \propto e^{-t/\tau_d}, \quad i \neq j\end{equation}
		where $\tau_d$ is the characteristic decoherence timescale. Over time we see $I \to 0$. The interference terms are suppressed, and
		\begin{equation}\rho_S \approx |c_1|^2|S_1\rangle \langle S_1| + |c_2|^2|S_2\rangle \langle S_2|\end{equation}
		Thus, any measurement on $S$, entangled with $E$, effectively ignores quantum interference between superposed states.
		
		Alternatively, decoherence can be understood through the decoherent histories framework by Gell-Mann and Hartle (1990).\footnote{\x{See also Wallace (2012, \S3.7) for an exposition.}} This framework doesn't require environmental interaction. A 'history' is a sequence of projection operators $\{P_1, P_2, \ldots, P_n\}$ at successive times $t_1, t_2, \ldots, t_n$. The chain operator $C_\alpha$ for a given history is:
		\begin{equation}
			C_\alpha = P_n U(t_n, t_{n-1}) P_{n-1} \cdots U(t_2, t_1) P_1.
		\end{equation}
		The probability for a history $\alpha$ is given by the decoherence functional $D(\alpha, \alpha')$:
		\begin{equation}
			D(\alpha, \alpha') = \text{Tr} \left( C_\alpha \rho_0 C_{\alpha'}^\dagger \right),
		\end{equation}
		where $\rho_0$ is the initial density matrix. The diagonal elements $D(\alpha, \alpha)$ provide the probability of the history $\alpha$ occurring. Histories must satisfy the decoherence condition to be physically meaningful, ensuring interference between different histories vanishes:
		\begin{equation}
			D(\alpha, \alpha') = 0 \quad \text{for} \quad \alpha \neq \alpha'.
		\end{equation}
		Dynamics is crucial since the unitary operators $U(t_{i+1}, t_i)$ govern the system's state evolution between projections. Projection operators must be compatible, forming a complete set of orthogonal projections at a time:
		\begin{equation}
			\sum_i P_i(t) = I,
		\end{equation}
		and
		\begin{equation}
			P_i(t) P_j(t) = \delta_{ij} P_i(t).
		\end{equation}
		These conditions ensure that the set of projections $\{P_i(t)\}$ at each time $t$ represents a mutually exclusive and collectively exhaustive set of alternatives for the system's state.
		
		Returning to the semiclassical time program: starting with the timeless wave-function, even if this wave-function is in an arbitrary superposition of states, decoherence is supposed to let us effectively ignore quantum interference and approximate the wave-function of our world as an eigenstate, eventually allowing us to derive semiclassical time. However, it seems clear that decoherence presumes temporal evolution by the Schrödinger equation on either approach to decoherence! (The first approach also raises a further question as to what the `environment' -- and that which induces decoherence -- of a subsystem is supposed to be, when the subsystem is the geometry and matter content of the universe.) Without time, why assume decoherence will occur?\footnote{See Chua and Callender (2021) for discussion.}
		
		One finds this tension in Kiefer's own account. On the one hand, he writes that 
		\begin{quote}
			A prerequisite [of decoherence in the semiclassical time program] is the validity of the semiclassical approximation for the global variables. This brings an approximate time parameter $t$ into play. (Kiefer 2007, 311)
		\end{quote} 
		On the other hand, he writes later that 
		\begin{quote}
			Since [decoherence] is a prerequisite for the derivation of the Schrödinger equation, one might even say that time (the WKB time parameter in the Schrödinger equation) arises from symmetry breaking [i.e decoherence]... Strictly speaking, \textit{the very concept of time makes sense only after decoherence has occurred}." (Kiefer 2007, 317--318, emphasis mine)
		\end{quote} But there is some tension here: on this interpretation, we needed time in the timeless formalism to get decoherence going, but we need decoherence in order to get time into the timeless picture. We cannot interpret step 1 in a way that is consistent with a fundamentally timeless ontology, and so we cannot get an explanation of how time emerges from no-time. \z{Note, in particular, that my claim, then, is not that Kiefer's framework is \textit{formally inconsistent}, but that under standard interpretations its physical coherence has not yet been fully secured. This is thus an invitation to defenders of the semiclassical approach to say more (see \S3.2).}
		
		Might we avoid this by providing some new physical interpretation of decoherence? To my knowledge, Castagnino and Laura (2000) are the only ones who posit something that might help, a sort of `primitive decoherence'. They write that \begin{quote}
			we could postulate (based on a obvious observational fact) that the quantum physics of the universe is such that it has a tendency towards a classical regime [decoherence]. This tendency would be the primitive concept in this case and time would be a derived concept... (9)
		\end{quote}
		However, \z{the worry is again one of physical coherence: why think that decoherence occurs even in a fundamentally timeless world, and even if there were no dynamics to begin with?} Now, I do think they are on the right track, insofar as the semiclassical time theorist must now provide a new interpretation of decoherence which doesn't appeal to time or dynamics (or an environment) like the standard stories above, on pains of circularity and physical incoherence. However, this postulate is at best a \textit{heuristic}, in order to motivate the search for an appropriate explanation and physical reason for why decoherence occurs even for timeless wave-functions. Even if we could swallow the ad-hoc nature of such a move, it is simply not clear yet just \textit{how} primitive decoherence is supposed to work, and what the physical motivation for such a move could be.\footnote{\z{Setting aside the worry of physical coherence, note that we can nonetheless do quite a lot of work as to what sort of formal features such a physical account of decoherence should satisfy (see Chataignier et al. (2025) and references therein for such calculations).}} 
		
		Without this story, it seems apparent to me that there is no physically satisfactory interpretation of decoherence that passes the test of physical coherence: on two plausible interpretations, the tool of decoherence appears to require time in order to justify its application, and hence its application in the timeless context appears to be simply \textit{incoherent}. A fortiori, the semiclassical time program also appears to be physically incoherent for now. 
		
		Crucially, note that we have not brought in any external considerations of metaphysics, e.g., mereological considerations; rather, the worry arises simply from considerations internal to the relevant physics e.g., an appropriate physical interpretation of the formal tool of decoherence employed in the semiclassical time framework. 

        \z{Before moving on, one might worry that my criticism goes too far.\footnote{I thank an anonymous reviewer for suggesting this worry.} In particular, the derivation of semiclassical time here is formally similar to that of Mott's (1931) derivation of the time-dependent Schr\"odinger equation from the time-\textit{in}dependent Schr\"odinger equation. If those formal moves worked in that context, why not in the context of semiclassical time? However, a central point of the analysis of physical coherence -- and a point we will return to in the next section -- is precisely that \textit{formally similar derivations can have quite different physical interpretations}. As Briggs and Rost (2001) point out in their refinement of Mott's derivation of the time-dependent Schr\"odinger equation from a time-independent one, in that context,
        \begin{quote}
            the starting point is the TISE for a closed, energy conserving, quantum object comprised of two parts, called the system and the environment. \textit{In the limit that the environment can be treated classically, it provides a time variable with which to monitor the remaining quantum system whose development, as viewed from the environment, is governed by the TDSE for the system alone.} This derivation shows explicitly that the origin of the classical time ... is due to coupling with the classical environment, and that the parametric derivative $\partial/\partial t$ arises from the transition of environment variables from quantum to classical behaviour. (694)
        \end{quote}
        Thus, note, crucially, that the environment must already be assumed to be \textit{temporal} in standard Mott-style derivations, in order for the subsystem to be describable by a time-dependent equation. However, in the case of semiclassical gravity, the starting point is \textit{precisely a timeless world}. The worry of physical incoherence therefore arises only in this case -- and not in standard Mott-style derivations -- precisely because we cannot assume the `heavier' gravitational `environment' to be temporal in nature to begin with, without abandoning the premise of the middle way that time fundamentally does not exist.}  
        
		\subsection{Interlude: Huggett and Th\'ebault}
		
		Recently, Huggett and Th\'ebault (2025) (\textbf{HT}) have argued against \textbf{CC}'s analysis that the semiclassical time program faces physical incoherence. Notably, they argue that the Born-Oppenheimer and WKB approximations, targets of \textbf{CC}'s analysis, don't necessarily require time for their interpretation and application. If so, it might turn out that the semiclassical time program does avoid physical incoherence after all contra \textbf{CC}. 
		
		The first and most important response is this: I am open to this possibility. After all, the point of developing this idea of physical coherence is to provide an operable and contingent standard with which one can check and test the viability of quantum gravity programs purporting to walk the middle way. If semiclassical time turns out to be physically coherent after all, we come out from the final analysis with (1) a plausible proposal for the middle way, and (2) a much clearer foundational understanding of how the middle way is supposed to work in the semiclassical time program. That is progress in my view. 
		
		With that said, I think that there are areas where their criticisms of \textbf{CC}, and defense of the semiclassical time program, merit more explanation. I'll focus on their discussion of the Born-Oppenheimer approximation since this was their main focus as well.  
		
		To begin, they argue (2025, 8--9) that, \textit{as a matter of fact}, the Born-Oppenheimer approximation in the ordinary molecular case already doesn't require an appeal to dynamics. Contra \textbf{CC} (2021, \S3.1), they claim that we don't need to appeal to the relative dynamics of `lighter' or `heavier' subsystems for applying the Born-Oppenheimer approximation (\textbf{HT}, 9). They conclude that ``the approximation requires neither time dependent dynamics nor ‘fixed’ nuclear positions for its application". Rather, ``this is a \textit{formal technique to construct approximate solutions to a partial differential equation} within a well controlled regime of validity." (emphasis mine) If so, then even the standard physical interpretation of the Born-Oppenheimer approximation might not need time, and so we shouldn't expect it to be physically incoherent in the timeless quantum gravity context.
		
		To motivate this claim, they provide an alternative story for applying this approximation by specifying three steps for applying it to the molecular case. First, one solves the eigenvector equation:
		\begin{equation}
			\left( \hat{T}_2 + \hat{W}(x_1) \right) \psi_n(x_1; x_2) = \lambda_n(x_1) \psi_n(x_1; x_2)
		\end{equation}
		where 1 is the `heavier' system and 2 is the `lighter' system in some sense, $T$ the kinetic energy, $W$ the potential, $\lambda$ the eigenvalues to be solved for. Interestingly, they interpret this equation as such:
		\begin{quote}
			This equation treats the ‘heavy’ subsystem as if it were fixed at a definite location $x_1$, and the ‘light’ subsystem as moving in the resulting potential $\hat{W}(x_1)$; for instance, the equation could describe how electrons would move for a certain classical configuration of nuclei. (8) 
		\end{quote}
		which seems to run counter to their central claim for the molecular case (in line with \textbf{CC}'s analysis). In a footnote, they cash out the `as if' which plays a crucial role: 
		\begin{quote}
			This step is typically given a heuristic gloss in  terms of treating the `heavy’ subsystem as if it were `fixed’ or `clamped’ at a definite location $x_1$, and the ‘light’ subsystem as moving in the resulting potential $\hat{W}(x_1)$. Physically, however, there is no sense in which the nuclei are literally fixed at points, since they are quantum objects. Rather the step should be understood as a formal move in an approximation scheme, and the ‘fixed’ description a heuristic gloss, which is ultimately both unphysical and unnecessary. (fn. 12)
		\end{quote}
		That is, while the earlier quote is meant to be a heuristic gloss, this equation is ultimately just math -- a pure formal move (they make a similar point for WKB). But without interpretation in terms of physical systems, what does the equation mean and what is it saying about those physical systems? Of course, the nuclei are not literally fixed in space, so that cannot be the interpretation. How else, then, are we supposed to interpret the equation? And without an interpretation, how do we know what the equation is saying about any physical system in the world? I insist that a formal move must be given some physical interpretation, and is just as unphysical otherwise. For instance, one might say, following \textbf{CC}, that the `heavy' subsystem is moving so slowly relative to the `light' subsystem that we can treat the `heavy' subsystem as effectively or approximately not moving relative to the dynamical timescales of the `light' subsystem. But that's a manifestly dynamical story which I believe would not be satisfactory to \textbf{HT}, and would also run afoul of physical incoherence. More, then, needs to be said.
		
		The second step faces a similar worry. We solve (14) and obtain:
		\begin{equation}
			\Psi(x_1, x_2) = \sum_n c_n \theta_n(x_1) \psi_n(x_1; x_2)
		\end{equation}
		and make what they call the separation ansatz so that:
		\begin{equation}
			\Psi(x_1, x_2) \approx \theta_n(x_1) \psi_n(x_1; x_2).
		\end{equation}
		for some normalized coefficients $\theta_n(x_1)$. They motivate this ansatz by appealing to the following reasoning:
		\begin{quote}
			The rationale for this separation is that because of the separation of masses, there will be far more kinetic energy in the light subsystem than the heavy one, and $E \approx \lambda_n$.(8)
		\end{quote}
		They argue that ``the mass separation is rather proxy for the electronic energy level separation, so a property of solutions of the time-independent Schr\"odinger equation" (19), and hence not dependent on dynamical notions at all. But of course, one can ask what the electronic energy level separation -- specifically due to the large differences in \textit{kinetic} energy between the two subsystems -- amounts to for a physical system like an electron. It seems to me just another way of saying that the heavier system isn't moving much compared to the light subsystem and a dynamical notion after all. Again, of course, one could simply respond that this is just math; my response would be the same as before for step 1. 
		
		The third step follows from the second step, which allows us to make the adiabatic approximation:
		\begin{equation}
			\frac{\partial \psi_n (x_1; x_2)}{\partial x_1} \approx 0.
		\end{equation}
		Note what this means: variations of the `heavy' subsystem's position parameter leave the `light'  system's wave-function approximately unchanged, which then lets us solve for the coefficients $\theta_n(x_1)$. Though unsaid, the most natural interpretation of this equation seems to provide a justification for assuming the approximately `fixed' location of the heavier system: the `light' system's wave-function is essentially insensitive to \textit{motions} of the `heavy' system, i.e. changes to its positions. Of course, time doesn't enter explicitly here, but it seems to me obvious in the ordinary case that changes of positions occur under changes in time.  
		
		It thus seems to me that contrary to what they said, a physical interpretation of the approximation in the molecular case does implicitly demand dynamical reasoning, as \textbf{CC} argued. Their appeal to the formal nature of this approximation to justify its atemporal nature seems inadequate to me: as a matter of math, it is certainly atemporal just like Plato's Forms. But what we want to know is whether its interpretation in terms of actual physical systems requires time and dynamics.
		
		Now, moving to the timeless Wheeler-DeWitt case (\textbf{HT}, 10--11), they acknowledge that we cannot even justify the second step (and hence the third step) as presented above. Instead, they suggest the following recipe: apply step 1, apply step 3 \textit{as an ansatz}, apply the WKB approximation, then simply check by explicit calculations when step 3 holds and hence when ``the Born-Oppenheimer method is formally valid" (11). They don't, however, provide an interpretation -- not even heuristic ones! -- for what is going on when these approximations formally hold. 
		
		I've already noted that the most natural physical interpretation of the first step seems dynamical in nature. They make the same move here -- that this is merely a heuristic -- when they say that we apply the first step ``\textit{as if} the field were propagating in a fixed spatial geometry $\alpha$" (10, emphasis mine), where the `light' matter field is supposed to be contrasted against the `heavy' $\alpha$. Taken literally, this, of course, sounds like it requires some temporal or dynamical interpretation of both systems, running into physical incoherence. ``As if" is playing a big conceptual role here in motivating step 1 in some non-literal way, but they don't say much here. 
		
		Furthermore, again, I am not sure that appealing to formal ansatz alone can do the job of securing physical coherence. I fully agree that obtaining the formal validity of the Born-Oppenheimer approximation is an achievement, and a \textit{heuristic} (to borrow their words) for future work in securing the physical coherence of time's emergence in the semiclassical time program. But, calling back to the discussion of \S2, it cannot be the \textit{end} of the task: we must provide a physically meaningful story of what exactly is going on in the world as we move from the fundamental timeless ontology to a non-fundamental temporal one via these formal tools. A refusal to provide a physical interpretation of these formal tools doesn't prove that the story is physically coherent; it just leaves the story incomplete, leaving the physical meaning of the semiclassical time program unspecified. \x{Furthermore, the appeal to heuristics appears to trade on ambiguity: we are supposed to find the application of some formal tool justifiable or acceptable because of the heuristic interpretation of that formal tool, yet we are not supposed to take the heuristic interpretation too seriously. Why, then, should we take the acceptability of the formal tool's application too seriously?} The worry to be assuaged here is that crucial formal steps -- e.g. step 3's adiabaticity -- might turn out to sneak in temporal assumptions once we actually provide a physical interpretation, leading to physical incoherence. 
		
		Finally, note that they don't discuss decoherence at all. They note that the second step -- the separation ansatz -- cannot be justified in the Wheeler-DeWitt case in the usual way. This leads them to appeal to explicit calculations and formal validity as the standard of justification. However, what they don't discuss is Kiefer's own motivation for the separation ansatz via decoherence (2007, 168). Decoherence is, for Kiefer, a necessary step for the Born-Oppenheimer approximation's applicability in the quantum gravity context, and hence for walking the middle way in the semiclassical time program. Of course, it is also the one that is most obviously dynamical in nature as discussed in \textbf{CC} and the discussion above.  
		
		Ultimately, it's work to be done -- hard work! -- to settle the question of semiclassical time's physical coherence. Nevertheless, I think this discussion shows the fruitfulness of the criterion of physical coherence: it demands explanation and interpretation for the various formal moves made by proposals for the middle way, and scrutinizes where they might go wrong. This can lead to refinement of the original proposal, as \textbf{HT} has done. \x{The scrutiny may also open new lines of interpretation and inquiry; for instance, one way to extend \textbf{HT}'s defense of the Born-Oppenheimer strategy (though not something that they pursue, to my knowledge) is whether the idea of energy is necessarily tied to the idea of motion or stationarity over time. If not, we might be able to provide an interpretation of energy gaps -- used to justify the Born-Oppenheimer approximation -- that doesn't succumb to my above criticism about the use of kinetic energy in \textbf{HT}'s `heuristic' interpretation. In turn, this might require an account of energy that is none of the classical forms of energy such as kinetic energy. There's more work to be done in clarifying the physical coherence of semiclassical time.\footnote{See also Bamonti et al.'s (2025) criticism of other features of \textbf{HT}'s physical story, which likewise grants much of the formal machinery.}   Either way, the demand for physical coherence is fruitful.} Either semiclassical time turns out to be physically incoherent, or not. In both cases, we would have gained some new insights into quantum gravity and the middle way.
		
		\subsection{Thermal time}
		
		Let us now apply the search for physical (in)coherence to another proposal for the middle way: the thermal time hypothesis (TTH), as developed by Connes and Rovelli (1994). The core idea, too, is simple: instead of the semiclassical level, time emerges from the timeless world at the non-fundamental thermodynamic level. 
		
		The proposal itself is cashed out in technical terms via what are known as von Neumann algebras and the Kubo-Martin-Schwinger (KMS) condition, but I'll summarize the conceptual core of the proposal in simpler terms.\footnote{See Swanson (2021) or Chua (2025) for discussion.} Starting again with $(a_{\text{TT}})$ the Wheeler-DeWitt wave-functional which is $(a'_{\text{TT}})$ interpreted as representing a timeless world, we are asked to consider spatial or matter distributions in the timeless world. Then, as we zoom out and consider the world in non-fundamental terms, we claim that some objects can be seen to be in $(b_{\text{TT}})$ formal states satisfying the KMS condition, which are $(b'_{\text{TT}})$ interpreted to be states of thermal equilibrium. Then $(c_{\text{TT}})$ one defines a one-parameter automorphism group and its generator that can be $(c'_{\text{TT}})$ interpreted as a time-translation group and the time-translation generator, i.e. the Hamiltonian. In short, time emerges from thermodynamics. 
		
		Chua (2025) discusses several worries with this approach at great length, but, again, due to space constraints I'll only consider the physical coherence of one crucial step for the program. To derive thermal time from no-time, one must first define states in thermal equilibrium on the timeless ontology. However, physical incoherence looms: the formal tool of KMS states interpreted as states of thermal equilibrium appears to be one that requires time in order to be applied. 
		
		\subsubsection{The time in equilibrium}
		
		Emch and Liu (2002, 355) discuss the standard properties of the equilibrium concept: systems in equilibrium are stationary, passive, stable, and so on. However, the crucial point here for our purposes is that these properties are \textit{all inextricably steeped in time}. As Callen puts it emphatically in his well-known textbook: ``in all systems there is a tendency to evolve toward states in which the properties are determined by intrinsic factors and not by previously applied external influences." These are the equilibrium states, which are ``by definition, \textit{time independent}" (1985, 13) such that ``the properties of the system must be independent of the past history". (1985, 14) But, as with decoherence, if we need time in order to define equilibrium to begin with, then, again, the story of how time emerges from a timeless world via thermodynamics cannot succeed and the thermal time program appears to be physically incoherent in this timeless context.
		
		As with Castagnino and Laura's proposal for primitive decoherence, one finds essentially the same move made in the context of thermal time. As Paetz (2010, \S7.6) argues, for TTH to succeed, we need an intrinsic definition of equilibrium, one that doesn't refer to time. To my knowledge, Rovelli (1993) is the only author who provides a `timeless’ definition of equilibrium. 
		
		In ordinary statistical mechanics for a system $S$ with phase space coordinates $p, q$ in thermodynamic equilibrium,\footnote{Rovelli's proposal is supposed to extend beyond the instantaneous phase space of classical statistical mechanics into generalized phase spaces (compatible with the canonical approach to quantum gravity) where time is demoted, from a privileged, global, coordinate defining $p$ and $q$ at a time, to merely one of the many phase space coordinates.} the interaction Hamiltonian approximately vanishes when separating a small \x{subsystem} $S'$ with coordinates $p', q'$ from the much larger system $S''$ with coordinates $p'', q''$. Thus, $S$'s statistical state $\rho_S$ factorizes in terms of the statistical states associated with \x{subsystems} $S', S''$:
		\begin{equation}
			\rho_S(p, q) = \rho_{S'}(p', q') \rho_{S''}(p'', q'')
		\end{equation}
		This condition of \textit{factorizability} represents the statistical independence of one sub-system's statistical state from the other. One way to interpret this statistical independence is as representing a system being in equilibrium with itself by representing its parts (i.e. subsystems) as being in \textit{relative equilibrium with each other}. If these subsystems are in relative equilibrium, their thermodynamic properties won't change with respect to each other, and so their statistical states are independent of each other and factorize. Though usually a consequence of thermodynamic equilibrium, Rovelli elevates (18) to a \textit{definition} of equilibrium: ``we shall refer to equilibrium as a situation in which every small but still macroscopic component of the system is in equilibrium, in the usual sense, with the rest of the system." (1993, 1558--1559) Hence, if we can find systems whose parts are in relative equilibrium with each other (i.e. systems satisfying factorizability for some generalized coordinates $p, q$), then those systems are in equilibrium simpliciter. If this works, then even in the fundamentally timeless world described by the Wheeler-DeWitt equation, regions of these worlds may be ascribed thermodynamic equilibrium states, with which to define thermal time.
		
		However, Chua (2025) argues that this new definition doesn't seem capable of picking out all and only the systems in equilibrium. Consider a precisified version of Rovelli's definition per Chua. Firstly, a system is in equilibrium if and only if \textit{every} subsystem is in \textit{relative equilibrium} with the rest of the system, viz. $S'$ and $S''$ are in relative equilibrium for all choices of $S'$ and $S''$ such that $S'$ and $S''$ are still macroscopic \x{subsystems} and $S'$ is significantly smaller than $S''$. Secondly, two subsystems $S'$ and $S''$ are in relative equilibrium if and only if (18) holds. Chua argues that this definition is neither necessary nor sufficient for picking out all and only the systems in thermodynamic equilibrium, and hence is a physically inadequate definition of thermodynamic equilibrium. 
		
		Firstly, defining equilibrium in terms of relative equilibrium for all choices of $S'$ and $S''$ is too stringent. While a system would indeed be in equilibrium if each subsystem were in relative equilibrium with the rest of the system, this condition is not necessary and thus cannot define equilibrium. Even in non-generally covariant contexts, typical systems known to be in equilibrium don't meet the relative equilibrium criterion for all subsystem choices. Rovelli’s requirement is simply that $ S' $ and $ S'' $ are macroscopic subsystems, with $ S' $ being much smaller than $ S'' $. However, as Chua shows, without further constraints, there will always be possible gerrymandered `Maxwell's demon' partitions of the system into two subsystems: a small, disconnected collection of subsystems containing all the faster particles with higher momentum ($ S_{fast} $) and a much larger \x{subsystem} containing the slower particles with lower momentum ($ S_{slow} $). This type of partitioning implies that $ S_{fast} $ is at a much higher temperature than $ S_{slow} $ since it has higher mean kinetic energy. Consequently, relative equilibrium doesn't obtain for such partitions, even though the system overall is in equilibrium.
		
		Secondly, mere statistical independence, i.e., the factorization condition (18) \textit{alone}, is insufficient to determine if two subsystems are in relative equilibrium. Consider a simple case with two boxes, $ B_1 $ and $ B_2 $, where $ B_2 $ is much larger than $ B_1 $. These boxes are thermally insulated, electromagnetically shielded, and contain air at different temperatures. We can describe the joint system of $ B_1 $ and $ B_2 $ with a statistical state $ \rho_B $, which might be factorizable into $ \rho_{B_1} $ and $ \rho_{B_2} $, assuming perfect thermal insulation and mirrors preventing radiation transmission. Even if the factorization holds for some regimes, this doesn't mean the two boxes are in relative equilibrium since they are at different temperatures. Thus, the proposed 'timeless' definition of relative equilibrium based on factorization doesn't suffice to characterize the actual thermodynamic relationship between the boxes.
		
		Finally, the issue with using factorizability to characterize relative equilibrium becomes more pronounced when we consider generalized phase spaces and the problem of time, where we cannot restrict our considerations to systems and states at a single moment. For instance, consider a system $ S $ undergoing a probabilistic process where, at each time-step $ \tau $, the state of the system is either 1 or 0 with some probability, independent of past and future outcomes. If we partition the entire sequence of states across time (e.g. $10001010$) into sub-sequences (e.g. $1000$ and $1010$), these 'subsystems' will factorize, satisfying the timeless definition of relative equilibrium. However, it's evident that these subsystems are \textit{not} in relative thermal equilibrium; they are merely probabilistically independent of each other. Thus, the factorization condition fails to capture the physical meaning of thermal equilibrium in this context.
		
		Chua notes that these issues go away if we took (18) to hold \textit{over time} and allowed the subsystems to \textit{interact}, viz. if the definition were justified by appeal to background dynamics.\footnote{\x{Similarly, one might think that the counterexample provided is just one particular configuration, and that the \textit{average} momenta differences -- and hence temperature differences -- between $S$ and $S'$ vanish. However, we must be careful here. Firstly, averages appeal to ensembles or probabilistic distributions, and the standard ensembles (such as the Gibbs state) are explicitly defined in terms of time-invariance and thermodynamic coupling, e.g., with a large environment such as a heat bath. It is unclear how these apply in the timeless context, again bringing us back to the problem of physical coherence. Secondly, these averages are phase averages, and the orthodox statistical mechanical explanation for their relationship to actual systems is in terms of their equivalence to time averages given certain dynamical assumptions, e.g. ergodicity. Again, dynamics is required for us to connect phase averages to actual systems.}} Then the first worry vanishes since $S_{fast}$ quickly loses energy to $S_{slow}$ and $S$ equilibrates over time. Likewise, the second worry is dissolved since non-interacting subsystems would not be said to be in relative equilibrium, since they don't interact. Similarly, the different temporal parts of the probabilistic process never interact with each other and hence are not appropriate entities to be in relative equilibrium. However, these moves require time and dynamics.\footnote{Incidentally, this returns us to Landau and Lifshitz's original physical justification for (18): when we're allowed to take a system to be comprised of \textit{quasi-closed} systems -- when these subsystems \textit{interact weakly} with each other. (1980, Ch. 1, $\S2$).}
		
		Again, physical incoherence looms: \textbf{TTH} cannot take off in a timeless context because we seem to require time to define thermal equilibrium, but we require thermal equilibrium to define time. Thus, even if the math $(a_{\text{TT}})$-$(c_{\text{TT}})$ is formally consistent, we cannot pick out states with the right sort of physical interpretation -- equilibrium states -- from the fundamentally timeless ontology, with which the middle way may coherently proceed.\footnote{Some work in the right direction, in trying to discuss thermodynamics without time, can be found in Vidotto (2025).}

		\section{Concluding remarks}
		
		I have proposed a standard of physical coherence for walking the middle way: provide a physically consistent story that can walk us from fundamental timeless ontology to non-fundamental temporal ontology, with a clear physical meaning to each step of the derivations necessary to bring us there. In turn, I have suggested how two proposals for the middle way may run into worries of physical incoherence. 
		
		I distinguish my standard of physical coherence from Baron, Miller and Tallant's (2022) standard of metaphysical coherence, or some such analysis. While the evaluation of metaphysical incoherence relies on metaphysical accounts of emergence and rules out spacetime emergence in an `all-or-nothing' fashion, an analysis of physical coherence requires careful and contextual study of the physical standards associated with each procedure being used, the physical interpretations for such procedures, and whether there exists a consistent interpretation from no-time to time (as I've shown here). This is not to say that metaphysical coherence is dispensable, or inadequate. Rather, physical coherence can do a different kind of work. In both cases of semiclassical and thermal time, it is the analysis of physical coherence that pinpoints specific implicitly temporal assumptions of the respective derivations that generated physical incoherence. For the former it was in the assumptions of the semiclassical approximations and decoherence, and for the latter it was in the assumptions of thermal equilibrium. This highlights room for new work for physicists seeking to defend these programs: provide arguments and new interpretations for why these tools may nevertheless apply despite the failure of standard interpretations. In this way, demands for physical coherence can do work that analyses of metaphysical emergence never could, being operative at a more local, program-specific level. The analysis of physical coherence thus leaves open the possibility that some programs in quantum gravity may pass the test, while others might not. For instance, proponents of semiclassical time could answer (or dissolve) my worries in \S3.2, which can bolster its physical coherence. This puts my standard in touch with physical practice, by providing a standard that is operable and applicable to different quantum gravity programs in order to assess their viability. It can also -- as the discussion of semiclassical time suggests -- motivate work on improving the conceptual foundations of extant programs by providing and clarifying the physical meaning of the formal moves made in search of time from no-time.

        \section*{Acknowledgments}

        I would like to thank Craig Callender, Yichen Luo, Sai Ying Ng, Ray Pedersen, Shelly Yiran Shi, the audience at the Society for the Metaphysics of Science Meeting 2023 in Halifax, and the audience at the European Philosophy of Science Association’s 2023 biennial conference, for their helpful comments and suggestions. I also want to thank two anonymous reviewers for their suggestions and comments.

		\bibliographystyle{plainnat}
		\bibliography{bib.bib}
		\nocite{*}
		
	\end{document}